\documentclass[11pt,a4paper]{article}
\pdfoutput=1  
\usepackage{amsmath,amsfonts,amssymb,amsbsy}
\usepackage{mathrsfs,latexsym}
\usepackage{graphicx}
\usepackage{color}
\usepackage{booktabs}
\usepackage{multirow}
\usepackage{multicol}
\usepackage{subfig}
\usepackage{alpha}
\usepackage{macros_alpha}
\usebiblio{latticen}

\renewcommand{\strange}{\mathrm{strange}}
\newcommand{\ee}{{\mathrm e}}
\newcommand{\bracket}[3]{{\left\langle#1\left|#2\right|#3\right\rangle}}
\newcommand{\expectationvalue}[1]{{\left\langle#1\right\rangle}}
\newcommand{\PCAC}{{\mathrm{PCAC}}}
\newcommand{\fa}{{f_\mathrm A}}
\newcommand{\fp}{{f_\mathrm P}}

\begin{document}

\preprintno{%
TCDMATH 15-01\\
MS-TP-15-02\\
CP3-Origins-2015-001 DNRF90\\
DIAS-2015-1
}

\title{%
Non-perturbative improvement of the axial current in $\nf = 3$ lattice QCD 
with Wilson fermions and tree-level improved gauge action 
}

\collaboration{\includegraphics[width=2.8cm]{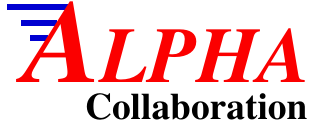}}

\author[trin]{John~Bulava}
\author[ode,esp]{Michele~Della~Morte}
\author[wwu]{Jochen~Heitger}
\author[wwu]{Christian~Wittemeier}

\address[trin]{School~of~Mathematics, Trinity~College, Dublin~2, Ireland}
\address[ode]{CP$^3$-Origins \& Danish IAS, University of Southern Denmark, Campusvej 55, 5230 Odense M, Denmark}
\address[esp]{IFIC and CSIC, Calle Catedr\'atico Jos\'e Beltran 2, 46980~Paterna, Valencia, Spain}
\address[wwu]{Institut~f\"ur~Theoretische~Physik, Universit\"at~M\"unster, Wilhelm-Klemm-Str.~9, 48149~M\"unster, Germany}

\begin{abstract}
The coefficient $c_{\mathrm{A}}$ required for 
$\mathrm{O}(a)$~improvement of the axial current in lattice QCD with 
$\nf=3$ flavors of Wilson fermions and the tree-level Symanzik-improved
gauge action is determined non-perturbatively. 
The standard improvement
condition using Schr\"{o}dinger functional boundary conditions is 
employed at constant physics for a range of 
couplings relevant for simulations at lattice spacings
of $\approx 0.09\,\Fm$ and below.
We define the improvement condition projected onto the zero 
topological charge sector of the theory, in order to avoid
the problem of possibly insufficient tunneling between 
topological sectors in our simulations at the smallest bare 
coupling.
An interpolation formula for $c_{\mathrm{A}}(g_0^2)$ is provided 
together with our final results.
\end{abstract}


\maketitle 

\section{Introduction} 

Wilson fermions~\cite{Wilson} are 
an attractive and popular fermion 
discretization for lattice QCD simulations. 
However, while the addition of the Wilson term lifts unwanted fermion 
`doubler' modes, it also leads to $\mathrm{O}(a)$ cutoff effects.
At small $a$, these cutoff  
effects are described by the (continuum) Symanzik effective 
theory~\cite{impr:Sym1, impr:Sym2}, and the procedure of their
systematic removal is called `$\mathrm{O}(a)$~improvement'.  

In order to eliminate these $\mathrm{O}(a)$ cutoff effects, a single 
dimension-five `clover' term must be added to the Lagrangian~\cite{impr:SW} 
(with coefficient $\csw$), 
and additional counterterms have to be added to local operators. 
To this end, the (unrenormalized) improved axial current is given by 
\begin{align}\label{e:aimp}
	&(A_{\mathrm{I}})^{a}_{\mu} = A^{a}_{\mu}(x) + a\,c_{\mathrm{A}} \frac{1}{2}
	(\partial_{\mu} + \partial_{\mu}^{*})P^{a}(x), 
	\\
	&A^{a}_{\mu}(x) = \bar{\psi}(x) T^{a} \gamma_{\mu}\gamma_5\psi(x), 
	\qquad P^{a} = \bar{\psi}(x) T^{a} \gamma_5 \psi(x), 
\end{align}
where $T^{a}$ is an $\mathrm{SU}(\nf)$ generator acting in flavor space 
and $\partial_{\mu}$, $\partial_{\mu}^{*}$ are the lattice forward and 
backward derivatives, respectively. 
Therefore, in order to improve matrix elements of the axial current, the 
coefficient $\ca$ must be specified as well as $\csw$. The 
non-perturbative determination 
of $\ca$ for a range of bare couplings is the subject of this work. 

Matrix elements of the axial current are of particular importance,
since they enter the computation of pseudoscalar meson decay constants and 
quark masses. 
These quantities are not only of great phenomenological interest in 
their own right, but it has also been demonstrated~\cite{lambda:nf2} that the 
kaon decay constant $\fK$ provides 
a precise determination of the lattice scale in physical units, 
thus affecting all dimensionful quantities. 
For instance, in the two-flavor theory and at lattice spacings of 
$\approx 0.1\,\Fm$, the $\ca$-term discussed above typically contributes 
$\approx10-15\%$ to pseudoscalar meson decay constants and quark 
masses~\cite{impr:ca_nf2,Kaneko:2007wh}. 

In this work we consider lattice QCD with $\nf = 3$ mass-degenerate flavors 
of Wilson fermions and the tree-level Symanzik-improved gauge 
action~\cite{Luscher:1984xn}. This gauge action is demonstrably preferable in
the pure gauge theory, where its cutoff effects are smaller than other standard 
actions~\cite{silvia:universality}. 
In preparation for dynamical simulations in this setup 
(see~\cite{Bruno:2014jqa} for a first report), 
the parameter $\csw$ 
multiplying the dimension-five `clover' term in the action has been tuned 
non-perturbatively~\cite{Bulava:2013cta}. 
Like $\csw$, the improvement coefficient $\ca$ has been determined in 
perturbation theory at 1-loop in Ref.~\cite{impr:csw_iwa_pert}. 
However, previous non-perturbative determinations of 
$\ca$~\cite{impr:ca_nf2,Kaneko:2007wh} deviate roughly $300-400\%$ from 1-loop 
perturbation theory at the largest bare couplings, so that a non-perturbative 
determination is required. 

To determine $\ca$ non-perturbatively, we employ the (by now) standard 
improvement condition for 
dynamical fermions~\cite{impr:ca_nf2, Kaneko:2007wh}, which imposes 
the PCAC relation at constant physics to ensure a removal of $\rmO(a)$ effects
in on-shell quantities and, at the same time, a smooth behavior of vanishing 
$\mathrm{O}(a^2)$ effects as the bare coupling is varied. However, for technical
reasons related to the topology freezing of our simulations at the finest 
lattice spacing, we project the necessary correlation functions onto the 
trivial topological sector. 
The main result of this work is the interpolation formula for $\ca(g_0^2)$ 
given in Eq.~(\ref{e:final}), with the coefficients of 
Eq.~(\ref{e:par_final}), which is valid for lattice spacings of 
$a\approx 0.09\,\Fm$ and below.
A statistical error of $4\%$ should be assigned 
to this formula
near the largest simulated bare couplings, whereas $7\%$ 
is more appropriate at the finest lattice spacing.  

We detail the 
improvement condition as well as the projection onto the trivial 
topological
sector in Sect.~\ref{s:impr}, and our simulation setup is discussed in 
Sect.~\ref{s:sims}. Numerical results together with the final interpolation 
formula are presented 
in Sect.~\ref{s:results}, and conclusions are drawn in Sect.~\ref{s:concl}.

\section{Improvement condition}\label{s:impr}

We determine $c_{\mathrm A}$ via a variant of the improvement condition 
originally introduced in quenched QCD in~\cite{impr:pap3} and applied in its 
present form to the theory
with dynamical fermions first in~\cite{impr:ca_nf2} for the two-flavor case. 
We briefly review this 
condition here, adopting the notation of Ref.~\cite{impr:ca_nf2}.

Improvement conditions are typically based on imposing the PCAC relation,
which in its continuum form is an operator identity, at finite lattice spacing.
A consequence of this relation is 
that the PCAC quark mass, defined as 
\begin{align}
  m_\PCAC & = \frac{\bracket{\alpha}{\partial_\mu A_\mu^a(x)}{\beta}}{2\bracket{\alpha}{P^a(x)}{\beta}}, 
\end{align}
is independent of the space-time position $x$ as well as the external states 
$\alpha$ and $\beta$. 
It is this property of $m_\PCAC$, required to hold on the
lattice up to $\Or(a^2)$ cutoff effects, which we exploit to 
determine $\ca$. 

Employing the above relation with the improved 
axial current of Eq.~(\ref{e:aimp}), we can decompose
\begin{gather}
  m_\PCAC = m(x;\alpha,\beta) = 
  r(x;\alpha,\beta)+a\ca\cdot s(x;\alpha,\beta), \\\nonumber\\
  \label{e:r_s_definition}
  r(x;\alpha,\beta) = \frac{\bracket{\alpha}{\frac{1}{2}(\partial_\mu+\partial^*_\mu)(A(x))^a_0}{\beta}}{2\bracket{\alpha}{P(x)^a}{\beta}}, \qquad
  s(x;\alpha,\beta) = \frac{\bracket{\alpha}{\partial_\mu\partial^*_\mu(P(x))^a}{\beta}}{2\bracket{\alpha}{P(x)^a}{\beta}}.
\end{gather}
As mentioned above, if the continuum form of the PCAC relation holds, 
$m(x;\alpha,\beta) = m$ is independent of $\alpha$, $\beta$ and $x$. 
If we demand this property and fix $x$, the quark mass $m$ can be eliminated 
by using two different pairs of external 
states $\alpha,\beta$ and $\gamma,\delta$ to obtain our definition of $\ca$: 
\begin{align}
  \label{e:ca_definition_general}
  \ca & = -\frac{1}{a}\cdot\frac{r(x;\alpha,\beta)-r(x;\gamma,\delta)}{s(x;\alpha,\beta)-s(x;\gamma,\delta)}.
\end{align}

In practice we employ a finite physical system size 
($L\approx 1.2\,\mathrm{fm}$) with Schr\"{o}dinger functional boundary 
conditions~\cite{SF:LNWW,SF:stefan1} in time 
and a periodic torus in space. Furthermore (using the standard notation), we 
employ a vanishing background gauge field ($\phi_i = \phi_i' = 0$) and 
periodic spatial boundary conditions for the fermion fields ($\theta_i = 0$). 
With the tree-level 
Symanzik-improved gauge action, there are several possibilities for 
implementing boundary $\mathrm{O}(a)$~improvement in the Schr\"{o}dinger 
functional. As 
in Ref.~\cite{Bulava:2013cta}, we resort to `choice B' of 
Ref.~\cite{impr:csw_iwa_pert}, which possesses the desirable property 
that the classical minimum of the action can be expressed analytically. Our 
boundary $\mathrm{O}(a)$~improvement is implemented at tree-level according to 
this choice. 
An additional boundary $\mathrm{O}(a)$~improvement 
possibility for this action in the Schr\"{o}dinger functional 
has been proposed recently in Ref.~\cite{Luscher:2014kea}.  

Correlation functions in the Schr\"{o}dinger functional setup may involve 
boundary quark fields, 
which are adopted here. We construct $r$ and $s$ 
defined above from the following correlation functions:  
\begin{gather}\label{e:corr1}
  \fa(x_0;\omega) = -\frac{a^3}{3L^6}\sum_{\vec x}\expectationvalue{A^a_0(x)O^a(\omega)}, \qquad
  \fp(x_0;\omega) = -\frac{a^3}{3L^6}\sum_{\vec x}\expectationvalue{P^a(x)O^a(\omega)}, \\\nonumber\\
  O^a(\omega) = a^6\sum_{\vec x,\vec y}\bar\zeta(\vec x)\cdot T^a\gamma_5\cdot\omega(\vec x-\vec y)\cdot\zeta(\vec y),
\end{gather}
where the pseudoscalar operator $O^{a}$ is composed of the boundary quark 
fields at $x_0 = 0$ ($\bar{\zeta}$~and~$\zeta$), while the `wavefunction' 
$\omega(\vec{x})$ will be discussed 
shortly. Defining the corresponding operator $O'^{a}(\omega')$ at $x_0 = T$, 
we also employ the boundary-to-boundary correlation function 
\begin{gather}\label{e:corr2}
  f_1(\omega',\omega) = -\frac{1}{3L^6}\expectationvalue{O'^a(\omega')O^a(\omega)}.
\end{gather}

As outlined before, we aim at probing the PCAC relation with two different 
pairs of external states. 
To achieve this, we construct approximate wavefunctions of the ground and
first excited state in the pseudoscalar channel, viz. 
\begin{align}
  \omega_{\pi^{(0)}} & \approx \sum_{i=1}^3\eta^{(0)}_i\omega_i, & \omega_{\pi^{(1)}} & \approx \sum_{i=1}^3\eta^{(1)}_i\omega_i,
\end{align}
where $\omega_{\pi^{(0)}}$ ($\omega_{\pi^{(1)}}$) is constructed to maximize the
overlap with the ground (first excited) state. 
These approximate wavefunctions are superpositions of trial wavefunctions 
with coefficients $\eta^{(i)}_j$, given by the eigenvectors of 
the matrix $f_{ij} = f_1(\omega'_i, \omega_j)$ corresponding to the largest 
and next-to-largest eigenvalues for $\omega_{\pi^{(0)}}$ and $\omega_{\pi^{(1)}}$, 
respectively. 
As our spatial trial (hydrogen-like) wavefunctions at the boundaries we take 
\begin{gather}
	\bar\omega_1(r) = \ee^{-r/a_0}, \qquad \bar\omega_2(r) = r\cdot\ee^{-r/a_0}, \qquad \bar\omega_3(r) = \ee^{-r/(2a_0)}, \\\nonumber\\
  \omega_i(\vec x) = N_i\sum_{\vec n\in\mathbb Z^3}\bar\omega_i(|\vec x-\vec nL|),
\end{gather}
where $N_i$ is a normalization factor chosen to ensure 
$a^3 \sum_{\vec{x}} \omega_i^2(\vec{x}) = 1$ and $a_0 = L/6$. 
Finally, our operational definition of $\ca$ now reads 
\begin{gather}
  \label{e:ca_definition}
	\ca(x_0) = -\frac{1}{a}\cdot\frac{r(x_0;i_0,\omega_{\pi^{(1)}})-r(x_0;i_0,\omega_{\pi^{(0)}})}{s(x_0;i_0,\omega_{\pi^{(1)}})-s(x_0;i_0, \omega_{\pi^{(0)}})}
	\equiv -\frac{1}{a} \frac{\Delta r(x_0)}{\Delta s(x_0)}
	, \\\nonumber\\
	r(x;i_0, \omega) = \frac{\frac{1}{2}(\partial_0+\partial^*_0)\fa(x_0;\omega)}{2\fp(x_0;\omega)}   , \qquad
  s(x;i_0, \omega) = \frac{\partial_0\partial^*_0\fp(x_0; \omega)}{2\fp(x_0;\omega)},
\end{gather}
which at the same time defines $\Delta r(x_0)$ and $\Delta s(x_0)$. 
In this way, the wavefunctions $\omega_{\pi^{(0)}}$ ($\omega_{\pi^{(1)}}$) 
determine the states $\beta$ ($\delta$), 
cf.~Eq.~(\ref{e:ca_definition_general}), while for both states $\alpha$ and 
$\gamma$ the plain Schr\"{o}dinger functional boundary state $i_0$ with 
vacuum quantum numbers is inserted.
The choice of $x_0$ will be discussed later. 

Unlike previous studies, we employ $T=3L/2$ lattices for the generation of
our dynamical gauge field ensembles,
with the intention of re-using them for a determination of the 
axial current renormalization constant $Z_{\rm A}$ according to the method of 
Ref.~\cite{DellaMorte:2008xb} (see~\cite{lat14:caza_nf3} for a preliminary 
report).
While in a quark mass independent improvement scheme as applied here
one ideally would impose the improvement condition of
Eq.~(\ref{e:ca_definition}) at zero quark mass, in practice it was found in 
Ref.~\cite{impr:ca_nf2} and is also confirmed by our data that $\ca$ is
rather insensitive to fairly small deviations from this constraint.
However, that is not the case for $Z_{\rm A}$, so in general 
we endeavor to tune the quark mass to values close to zero.

Like periodic temporal boundary conditions, Schr\"{o}dinger functional 
boundary conditions are `closed', and disconnected topological sectors 
emerge in the continuum limit. However, for small physical
volumes, non-trivial topological sectors receive a small weight in the 
partition sum.  
Unfortunately, to keep $\mathrm{O}(a^2)$ effects under control, and with an 
eye toward using our ensembles for $Z_{\rm A}$, our physical volume 
($L \approx 1.2\,\mathrm{fm}$) is large enough 
that non-trivial topological sectors are not completely suppressed. 
Whereas these 
sectors are sampled sufficiently at coarser lattice spacings, ensembles  
at our finest lattice spacing ($L/a=24$, $\beta=3.81$) are effectively 
frozen in the sector with topological charge $Q=0$. 

The issue of topology freezing in the Schr\"{o}dinger functional has been 
investigated recently in the pure gauge theory in Ref.~\cite{lat13:felix}.
There it was suggested that quantities projected to the zero topological 
sector have a smooth approach to the continuum limit. For the case at hand 
($\ca$ as well as $Z_{\mathrm{A}}$), quantities defined in this way differ from 
their conventional counterparts only by irrelevant cutoff effects.
Formally, we perform this projection for all observables of interest. 
Adopting the notation of Ref.~\cite{lat13:felix}, we define 
\begin{align}
\langle \mathcal{O} \rangle_{0} = \frac{\langle \mathcal{O} \, \delta_{Q,0}\rangle}{\langle \delta_{Q,0} \rangle}, 
\end{align}
where $\mathcal{O}$ is an arbitrary observable in the theory 
(such as the correlation functions in Eqs.~(\ref{e:corr1}) 
and~(\ref{e:corr2})), and the topological 
charge $Q$ is defined using the Wilson (gradient) flow~\cite{flow:ML,flow:LW} 
at flow time $t$ given by $\sqrt{8t}/L = c$ with
$c=0.35$. Since at finite lattice spacing the topological charge may take 
non-integer values, we define all configurations with $|Q| \le 0.5$ as the 
trivial topological sector. Thus (as in Ref.~\cite{lat13:felix}) we make the 
replacement $\delta_{Q,0} \rightarrow \theta(Q + 0.5)\theta(0.5-Q)$. 
For ease of notation we shall henceforth take $\ca$ to mean the one projected 
onto the trivial sector in this manner, with the exception of 
Tab.~\ref{t:res}, which directly compares results in all 
sectors (where available) with those restricted to $Q=0$. 
Let us anticipate already here that these two kinds of analyses yield
consistent results for $\ca$ as expected, because the Ward identities 
underlying the PCAC relation and thereby our improvement strategy hold in 
any topological sector.

An alternative to address topology 
freezing in the Schr\"{o}dinger functional has recently 
appeared~\cite{Luscher:2014kea}, namely the use of 
`half-open' boundary conditions. While not considered here, these boundary 
conditions may help 
the problem in future calculations provided an improvement condition which 
does not require boundary-to-boundary correlation functions is devised. 

\section{Simulation details}\label{s:sims}

Our simulations
are performed using Schr\"{o}dinger functional boundary conditions. We use 
the \texttt{openQCD} 
code\footnote{\url{http://luscher.web.cern.ch/luscher/openQCD/}}%
of Ref.~\cite{algo:openQCD}, which was also used for the simulations to
calculate $\csw$ in Ref.~\cite{Bulava:2013cta}.

The bare gauge couplings are chosen to approximately satisfy a constant 
physics condition, fixing $L \approx 1.2\,\mathrm{fm}$. 
In this way it is ensured that any O($a$) ambiguities in $\ca$ disappear 
smoothly toward the continuum limit.
For a thorough an more general discussion of the idea and virtues of
imposing improvement (and renormalization) conditions at constant physics,
see, e.g., Refs.~\cite{impr:babp_nf2,nara:rainer}.
Similarly to earlier work~\cite{impr:ca_nf2,Kaneko:2007wh},
we fix the physical volume by beginning with a particular pair of $g_0^2$ and
$L/a$ ($\beta =6/g_0^2 = 3.3$ at $L/a = 12$ in the present case) and choose 
the bare couplings for  
subsequent smaller lattice spacings according to the universal 2-loop 
$\beta$-function\footnote{%
Note that the non-universal 3-loop term of the $\beta$-function is not 
known for the tree-level Symanzik-improved gauge action.}. 
The range of lattice spacings covered in this way extends from 
$a\approx 0.09\,\Fm$ to $a\approx 0.045\,\Fm$.
As already mentioned, at each bare coupling we tune the 
bare quark mass so that the PCAC mass is kept approximately constant
and close to zero\footnote{%
Based on the experience from the two-flavor theory, for which the 
multiplicative quark mass renormalization factor only varies slowly with $a$,
we can safely neglect it for the tuning purposes here, too.}.
At several lattice spacings we confirm that 
our determination of $\ca$ is insensitive to variations of the (small) quark
mass.
\begin{table}
\centering
\renewcommand{\arraystretch}{1.25}
\setlength{\tabcolsep}{3pt}
\begin{tabular}{ccclccrrcc}
\toprule
$L^3\times T / a^4$ &&& $\beta$ && $\kappa$ & \#\,REP & \#\,MDU && ID   \\
\midrule
$12^3\times 17$     &&& 3.3     && 0.13652  & 10      & 10240   && A1k1 \\
		    &&&         && 0.13660  & 10      & 13048   && A1k2 \\
\hline  
$12^3\times 19$     &&& 3.3     && 0.13652  & 10      & 10468   && A2k1 \\
\hline  
$16^3\times 23$     &&& 3.512   && 0.13700  & 2       & 20480   && B1k1 \\
                    &&&         && 0.13703  & 1       & 8192    && B1k2 \\
                    &&&         && 0.13710  & 3       & 24560   && B1k3 \\
\hline 
$16^3\times 23$     &&& 3.47    && 0.13700  & 1       & 8176    && B2k1 \\
\hline
$20^3\times 29$     &&& 3.676   && 0.13680  & 1       & 7848    && C1k1 \\ 
                    &&&         && 0.13700  & 4       & 15232   && C1k2 \\
                    &&&         && 0.13719  & 4       & 15472   && C1k3 \\
\hline 
$24^3\times 35$     &&& 3.810   && 0.13712  & 7       & 15448   && D1k1 \\
\bottomrule
\end{tabular}
\caption{%
Summary of simulation parameters, number of replica and total number of 
molecular dynamics units of our gauge configuration ensembles labeled by
`ID'.  
}\label{t:sim}
\end{table}

Information about our ensembles, consisting of several replica per parameter
set in most cases, can be found in 
Tab.~\ref{t:sim}. Due to practical reasons discussed in the 
\texttt{openQCD} documentation\footnote{%
For this work we employ \texttt{openQCD} version 1.2. 
This issue has been corrected in the latest version (1.4).},
our lattices have temporal extents $T = 3L/2 - a$. 
Since this offset itself scales with $a$, one expects its influence on the
determination of $\ca$ to be of $\Or(a^2)$ and thus to be small;
still, we assess it on our coarsest lattice where it is largest. 
There we simulate $12^3\times17$ as well as $12^3\times19$ ensembles. 

Although it was found previously that small deviations from the constant 
physics 
condition have little effect on $\ca$~\cite{impr:ca_nf2}, we estimate this 
deviation by measuring the 
scale-dependent renormalized coupling $\bar{g}_{\rm GF}^2$, 
defined in Ref.~\cite{flow:FR}. 
Results for this coupling are shown in Tab.~\ref{t:res}.
We also test the dependence of $\ca$ on $L$ in physical units 
(and thereby on violations of the constant physics condition) directly by 
simulating an additional bare coupling at $L/a=16$.  

\begin{figure}
\centering
\includegraphics[width=\textwidth]{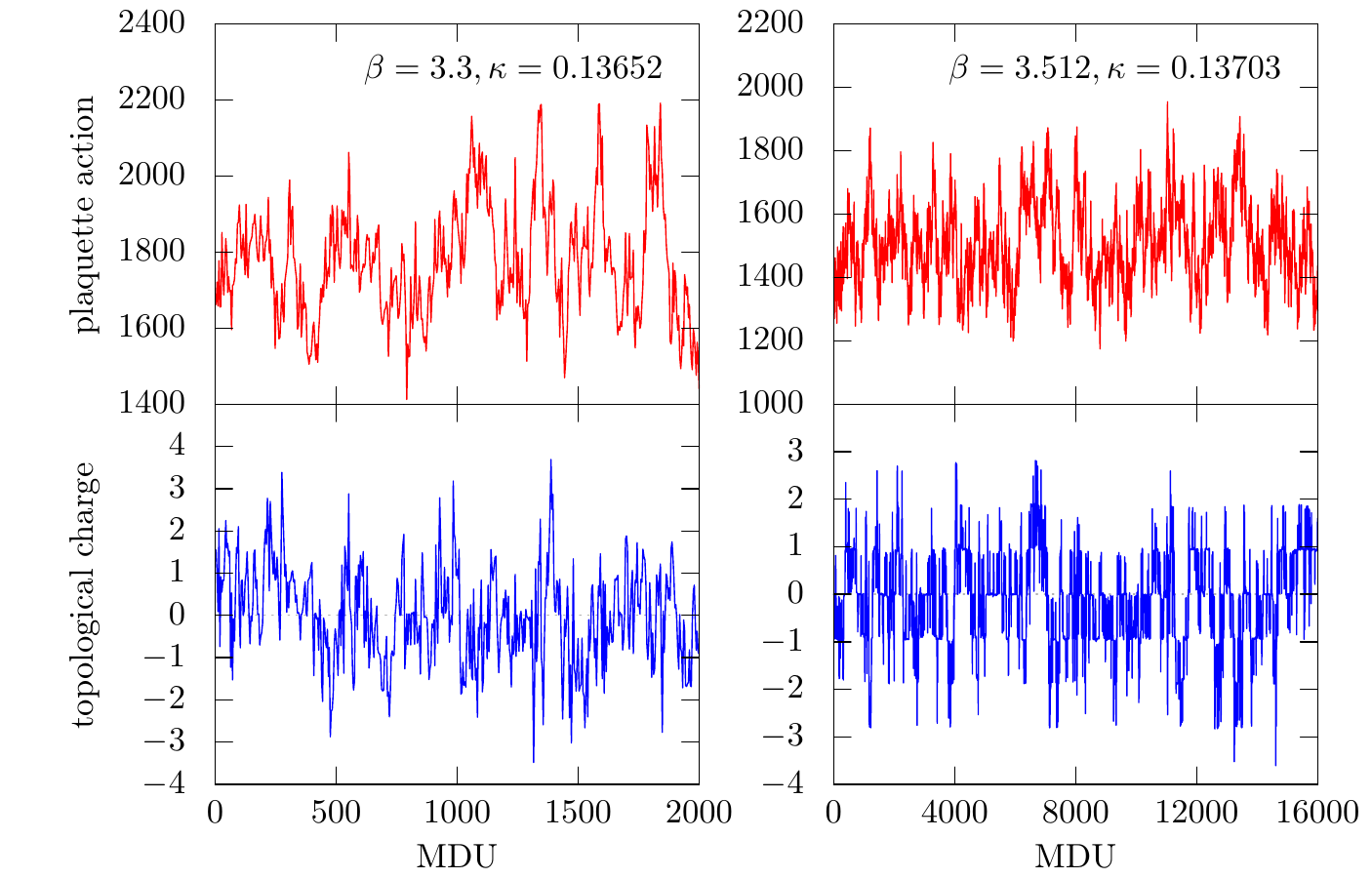} 
\includegraphics[width=\textwidth]{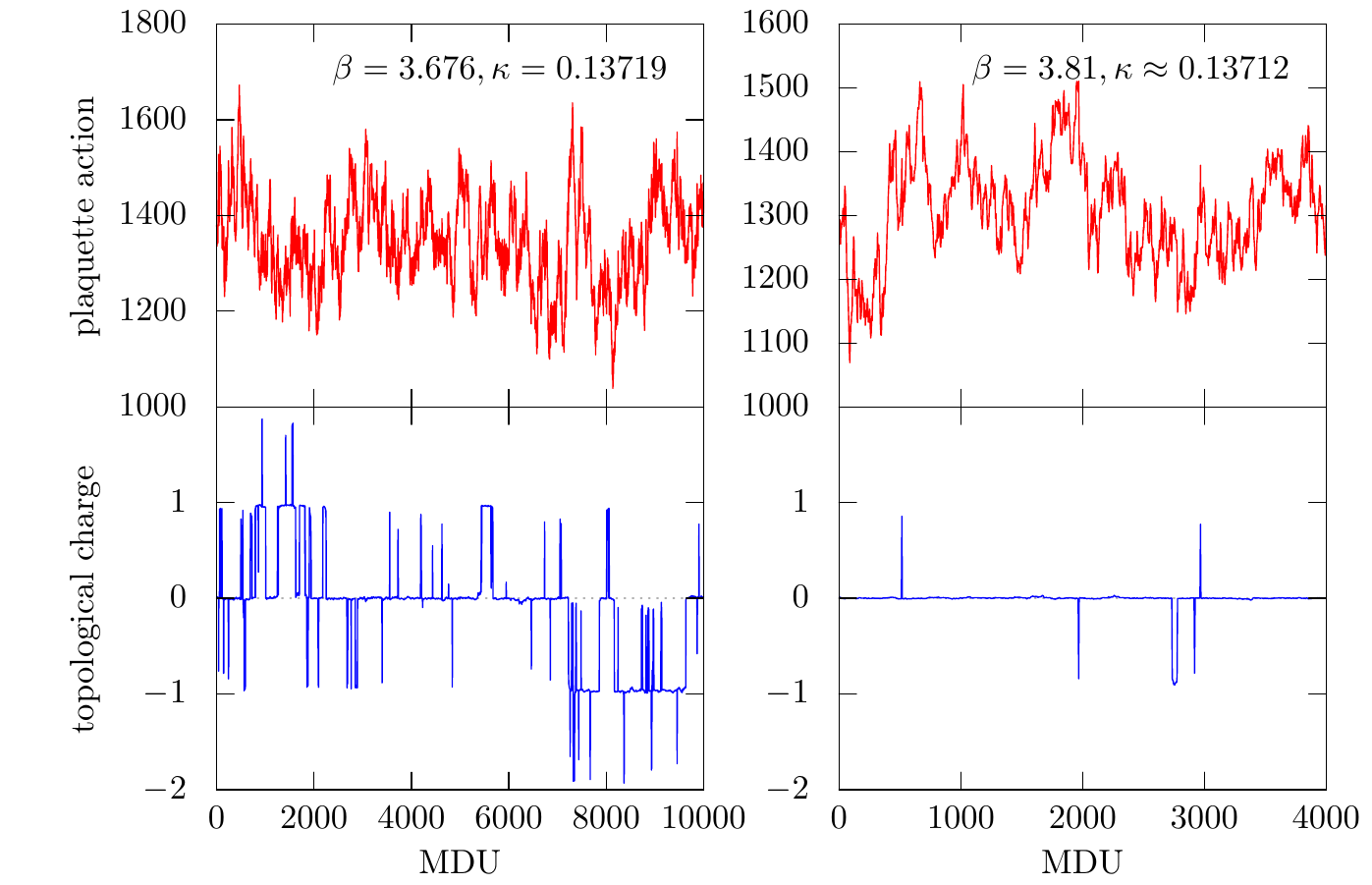} 
\caption{\label{f:freeze} Histories of the smoothed Wilson plaquette action
and the topological charge for single representative 
replica from each of the ensembles, which enter into the final 
analysis. Our inability to sufficiently sample all topological 
sectors at $\beta=3.81$ is evident.}  
\end{figure}
We now briefly summarize the simulation algorithm used for the generation of 
these gauge field ensembles.
While two of the (mass-degenerate) pseudo-fermion fields can be simulated in 
the usual way, the RHMC algorithm~\cite{algo:RHMC} is employed for the third. 
Even-odd preconditioning is used for all fermion determinants, 
whereas mass preconditioning~\cite{Hasenbusch:2002ai} with two additional 
pseudo-fermion fields is used for the degenerate doublet. 
We use a hierarchical integration 
scheme~\cite{Sexton:1992nu}, where the gauge force is integrated on the 
innermost level and the remaining fermion forces on the second level. 
For the $L/a = 16, 24$ and part of the $L/a=20$ lattices, 
the lowest poles of the RHMC are integrated 
on a third level. For the two inner levels, a fourth-order OMF 
integrator~\cite{Omelyan} is used, 
while the third level (when present) uses a second-order OMF integrator. 
A single step is used for the inner integrators, and the number of steps 
for the outer level is tuned to achieve an acceptance rate of $\approx 90\%$. 
For the coarsest $L/a=12$ lattice ensembles A1k2 and A2k1, we adopt (type I) 
twisted mass reweighting~\cite{lat08:filippo}. 
The conjugate gradient solver 
is employed for most fermion forces, while the multi-shift variant is 
typically used for most of the RHMC poles. 
For the lightest mass-preconditioned field and RHMC poles on the 
$L/a=16,24$ lattices, we employ the SAP-preconditioned 
GCR algorithm~\cite{Luscher:2003qa}.
\section{Results}\label{s:results}

On most of our ensembles, we measure 
the correlation functions defined in Eqs.~(\ref{e:corr1}) and~(\ref{e:corr2}) 
on every fourth trajectory of length $\tau=2\,\mathrm{MDU}$ so that the spacing 
between these measurements is $8\,\mathrm{MDU}$. Only on A1k2 and A2k1,
we use a measurement separation of $2\tau=4\,\mathrm{MDU}$. 
The total statistics for all the 
ensembles considered here are tabulated in Tab.~\ref{t:sim}. 

In addition to these correlation functions, we also measure `smoothed' 
gauge field observables obtained from the Wilson (gradient) 
flow~\cite{flow:ML,flow:LW}, 
which possess a well-defined continuum limit. These smoothed observables are 
useful in several ways. The smoothed gauge fields provide a renormalized 
definition of the topological charge, which we use to monitor the topology 
freezing discussed in Sect.~\ref{s:impr}. Even at lattice spacings 
where topology freezing is not a problem, the smoothed topological charge and 
action typically possess the largest observed autocorrelation times. 
Furthermore, the aforementioned renormalized (and $L$-dependent) coupling 
$\bar{g}_{\rm GF}^2$ of Ref.~\cite{flow:FR} is defined using the Wilson flow and 
may be used to monitor the deviation from the constant physics condition, 
as it is sensitive 
to the physical lattice size. Results for this coupling are given 
in Tab.~\ref{t:res}, too.

In order to monitor the autocorrelation times in our simulations, we examine 
these smoothed observables at a flow time $t$ given by $\sqrt{8t}/L = c$ with
$c=0.35$. For all simulations we find that integrated autocorrelation times 
of these observables satisfy the bound 
$\tau_{\mathrm{max}} \lesssim 200-250\,\mathrm{MDU}$, except for our $L/a=24$
simulations where the charge is frozen. The other smoothed observables turn
out to still possess autocorrelation times of comparable order of magnitude 
in these simulations.
The existence of the charge freezing at our finest lattice spacing is 
qualitatively illustrated in Fig.~\ref{f:freeze}. There it is seen 
that the autocorrelation time of the smoothed action remains under 
control, whereas the autocorrelation time of the topological charge increases 
significantly from the $L/a=20$ to the $L/a=24$ ensembles. 
We are practically unable to sufficiently sample all topological sectors at 
this finest lattice spacing, necessitating the restriction of our observables 
to the trivial sector.

\begin{table}
\centering
\renewcommand{\arraystretch}{1.25}
\setlength{\tabcolsep}{3pt}
\begin{tabular}{ccr@{.}lr@{.}lcccr@{.}l}
\toprule
ID && \multicolumn{2}{c}{$am_\PCAC$} & \multicolumn{2}{c}{$am_{\PCAC,0}$} & $\bar{g}_{\rm GF}^2$ & $\bar{g}_{\rm GF,0}^2$ & $\ca$ & \multicolumn{2}{c}{$c_{\mathrm{A},0}$} \\
\midrule
{\it A1k1} && $-0$&$0010(7) $ & $-0$&$0022(8) $ & 18.12(21) & 17.77(20) & $-0.0551(26)$ & $-{\it 0}$&${\it 0594(31)}$ \\
A1k2 && $-0$&$0086(6) $ & $-0$&$0100(8) $ & 16.95(13) & 16.62(15) & $-0.0557(19)$ & $-0$&$0552(24)$ \\
\hline  
A2k1 && $-0$&$0011(7) $ & $-0$&$0025(10)$ & 17.84(20) & 17.35(20) & $-0.0569(25)$ & $-0$&$0547(30)$ \\
\hline  
B1k1 && $0$&$0063(2)  $ & $0$&$0062(3)  $ & 16.49(13) & 16.44(14) & $-0.0365(11)$ & $-0$&$0348(15)$ \\
B1k2 && $0$&$0056(3)  $ & $0$&$0050(4)  $ & 16.85(20) & 16.57(23) & $-0.0381(16)$ & $-0$&$0334(29)$ \\
{\it B1k3} && $0$&$0022(2)  $ & $0$&$0016(3)  $ & 16.11(14) & 15.78(15) & $-0.0380(11)$ & $-{\it 0}$&${\it 0399(17)}$ \\
\hline 
B2k1 && $0$&$0041(4)  $ & $0$&$0036(5)  $ & 18.03(23) & 17.95(26) & $-0.0342(22)$ & $-0$&$0344(46)$ \\
\hline
C1k1 && $0$&$0138(1)  $ & $0$&$0137(2)  $ & 16.55(27) & 16.44(26) & $-0.0324(14)$ & $-0$&$0305(29)$ \\ 
C1k2 && $0$&$0066(2)  $ & $0$&$0065(3)  $ & 15.53(14) & 15.40(15) & $-0.0300(21)$ & $-0$&$0311(26)$ \\
{\it C1k3} && $-0$&$0005(1) $ & $-0$&$0006(2) $ & 14.64(13) & 14.41(16) & $-0.0281(14)$ & $-{\it 0}$&${\it 0291(14)}$ \\
\hline 
{\it D1k1} && n&q.            & $-0$&$00269(8)$ & n.q.      & 13.90(11) & n.q.          & $-{\it 0}$&${\it 0212(15)}$ \\ 
\bottomrule
\end{tabular}
\caption{%
Summary of results for $\ca$. 
The (unrenormalized) PCAC quark mass $am_\PCAC$ is computed from the 
correlation functions projected to the approximate ground state, 
using the 1-loop result for $\ca(g_0^2)$ from \cite{impr:csw_iwa_pert},
while $\bar{g}_{\rm GF}^2$ denotes the gradient (resp.\ Wilson) flow 
coupling mentioned in the text.
Recall that quantities with the explicit subscript label `0' here refer 
to results from the analysis restricted to the sector of vanishing 
topological charge, whereas in the text we loosely suppress the `0'.
Numbers for ensemble D1k1 ($L/a=24$) are not quoted (`n.q.') for the case
of covering all charge sectors in the partition sum, because our 
simulations are not able to sufficiently sample all the sectors and a 
reliable error estimation is thus not possible.  
Results from ensembles in italics enter into the final interpolation 
formula for  $\ca(g_0^2)$, Eq.~(\ref{e:final}).
}\label{t:res}
\end{table}

To provide further support for this projection, we have estimated the
expectation value $\langle\delta_{Q,0}\rangle$; it effectively corresponds
to the fraction of statistics, to which the restriction to $Q=0$ reduces the 
number of generated configurations.
For the ensembles with the smallest quark mass at given $L/a$,
\{A1k1, B1k3, C1k3, D1k1\}, we find $\langle\delta_{Q,0}\rangle$ $=$
$0.37(2)$ ($L/a=12$), $0.44(2)$ ($L/a=16$), $0.65(7)$ ($L/a=20$), 
$0.90(5)$ ($L/a=24$), and thereby to show large cutoff effects.
These kind of cutoff effects have been observed before in the large-volume, 
two-flavor theory in Ref.~\cite{Bruno:2014ova}, owing to the substantial 
suppression of the topological charge compared to the quenched approximation.
In addition, our $\langle\delta_{Q,0}\rangle$ at $m_\PCAC\approx 0$ does 
\emph{not} seem to smoothly (in the sense of monotonically \emph{and}
linearly or quadratically in $a$) approach~$1$ as $L/a$ increases, which is
the theoretical expectation for $\langle\delta_{Q,0}\rangle$ in the
large-volume continuum limit and at zero quark mass~\cite{Leutwyler:1992yt}.
Even though we are not in a large-volume situation with $m_\PCAC=0$ exactly, 
we interpret the encountered behavior of $\langle\delta_{Q,0}\rangle$ as a 
sampling problem of the algorithm, on top of the cutoff effects.
Therefore, we prefer to project our results to the $Q=0$ sector, which in the
end does not induce a noticeable difference in the final numbers for $\ca$, 
cf.~Tab.~\ref{t:res}.

The estimation of the statistical error on our measured quantities is
based on a single-elimination jackknife procedure, after first `binning' the 
data (concatenated from different replica) such that the size of each bin is 
$\gtrsim \tau_{\mathrm{max}}$, as well as on applying a 
full autocorrelation analysis according to Refs.~\cite{Wolff:2003sm,algo:csd}.
Both yield similar error estimates; our quoted final results in the $Q=0$
sector stem from the latter (without including any long-tail contributions
to the autocorrelation functions, which are negligible for the quantities
entering the $\ca$-analysis).

\begin{figure}[htb]
\centering
\includegraphics[width=0.625\textwidth]{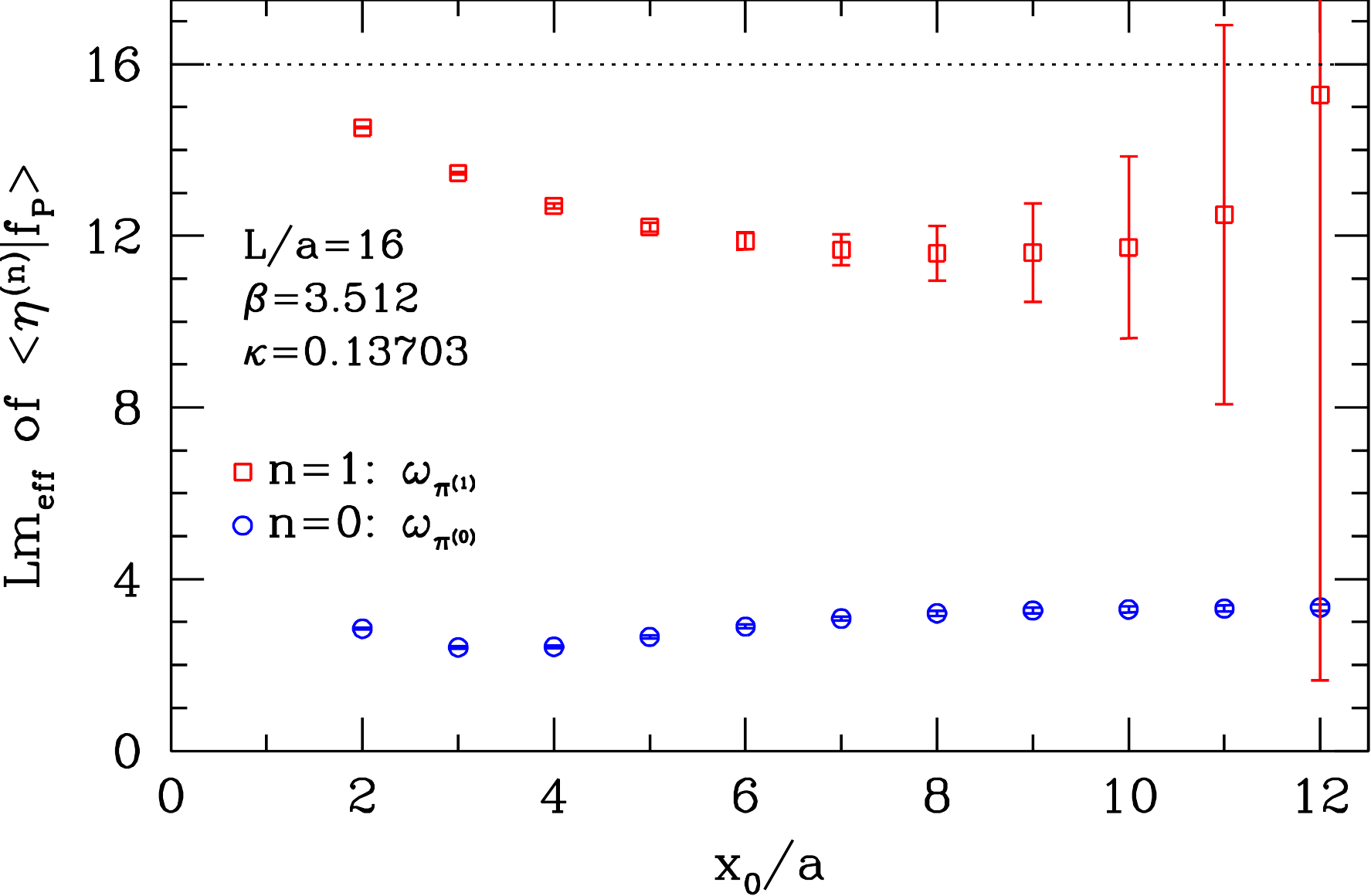}
\vskip0.5cm
\includegraphics[width=0.625\textwidth]{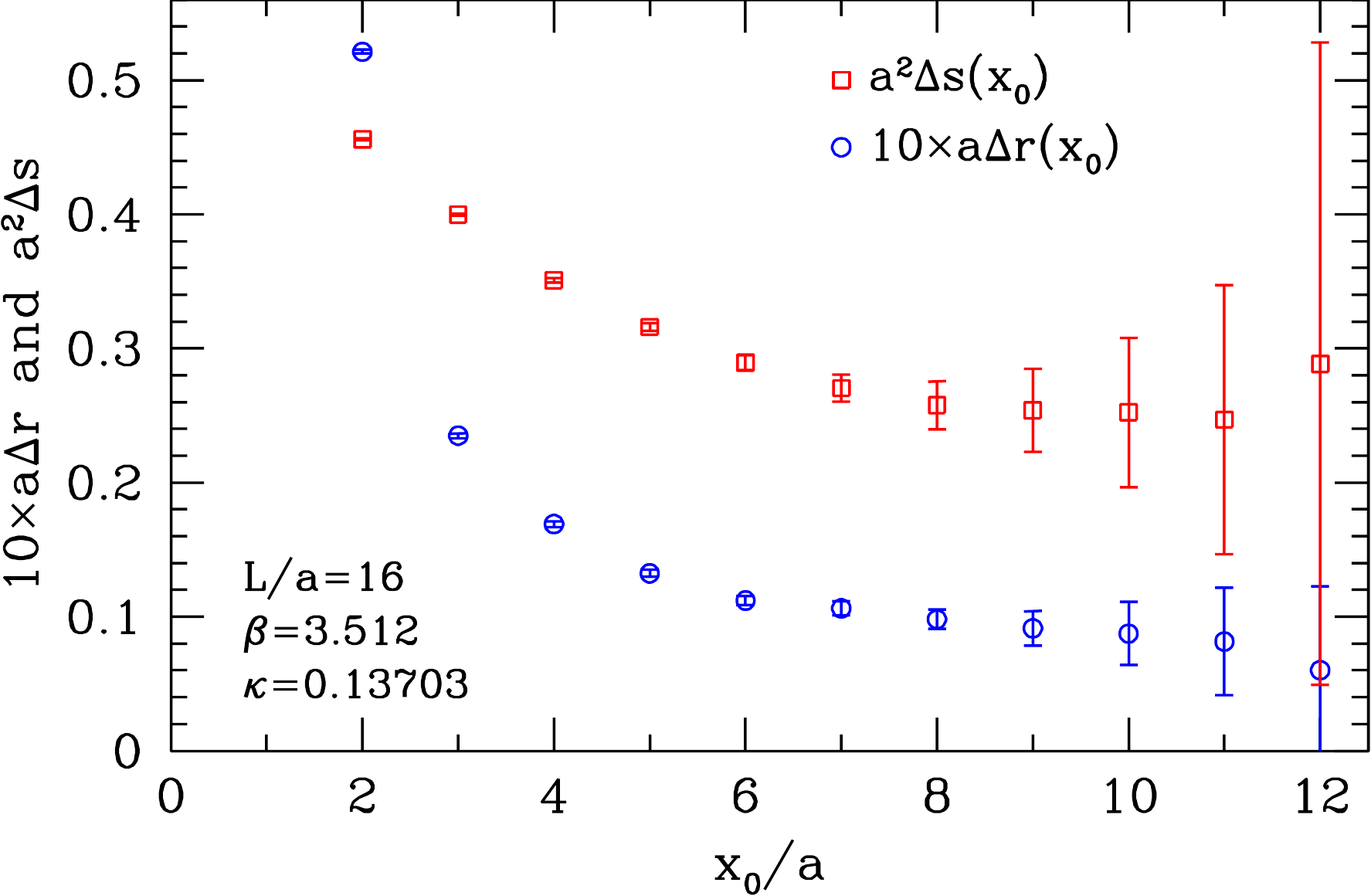}
\caption{\emph{Top:} Effective masses computed from $\fp$ with 
wavefunctions $\omega_{\pi^{(0)}}$ and $\omega_{\pi^{(1)}}$ 
(after projecting $\fp$ onto the approximate ground and first excited states) 
for the B1k2 ensemble.
The dotted horizontal line indicates the cutoff scale, $L\times a^{-1}=16$.
\emph{Bottom:} $\Delta s(x_0)$ and $\Delta r(x_0)$ for the same ensemble.}
\label{f:effective_masses}
\end{figure}
After measuring the correlation functions, we solve for the largest two 
eigenvectors of the matrix given by $f_1(\omega_i', \omega_j)$. 
These normalized eigenvectors have a 
well-defined continuum limit along our line of constant physics in 
parameter space, as long as the wavefunctions depend on physical scales only.
In fact, as we do not observe any significant lattice spacing dependence for
them, we fix these eigenvectors for once to the values calculated on the
B1k2 ensemble ($L/a=16$, $\beta=3.512$, $\kappa=0.13703$) and regard them 
as part of our choice of improvement condition.
For our setup we find:
$\eta^{(0)}=(0.5317(3),0.5977(1),0.6000(2))$ and 
$\eta^{(1)}=(0.843(5),-0.31(6),-0.44(6))$, which are similar to those of 
Refs.~\cite{impr:ca_nf2,Kaneko:2007wh}. 

To get an idea of the sensitivity of our method to $\ca$, we examine 
the effective masses of the correlation function $f_{\rm P}$, after taking the 
inner product with the eigenvectors 
for the (approximate) ground and first excited states. 
These are shown in 
Fig.~\ref{f:effective_masses} for the $L/a=16$ ensemble B1k2 of the previous
paragraph.
The distinctly seen signals display that indeed the eigenvectors 
effectively maximize the overlap with the ground and first excited states,
since these states are clearly separated up to $x_0\approx11a$.
As already noted in Ref.~\cite{impr:ca_nf2}, the energy of the first excited 
state is somewhat near the cutoff $a^{-1}$ though, and this may influence the 
way in which residual cutoff effects are modified.
Hence, residual $\mathrm{O}(a^2)$ effects may grow rapidly in 
smaller volumes~\cite{DellaMorte:2008xb}, which justifies our choice of an 
intermediate volume with $L \approx 1.2\,\mathrm{fm}$ to impose the 
improvement condition at constant physics.
Also shown in Fig.~\ref{f:effective_masses} are $\Delta r(x_0)$ and 
$\Delta s(x_0)$ (the latter being proportional to 
$m_{\pi^{(1)}}^2-m_{\pi^{(0)}}^2$ in case of exact ground and excited state
projections) from Eq.~(\ref{e:ca_definition}) for the same 
ensemble, further demonstrating our good sensitivity to $\ca$. 

\begin{figure}[htb]
\centering
\includegraphics[width=0.75\textwidth]{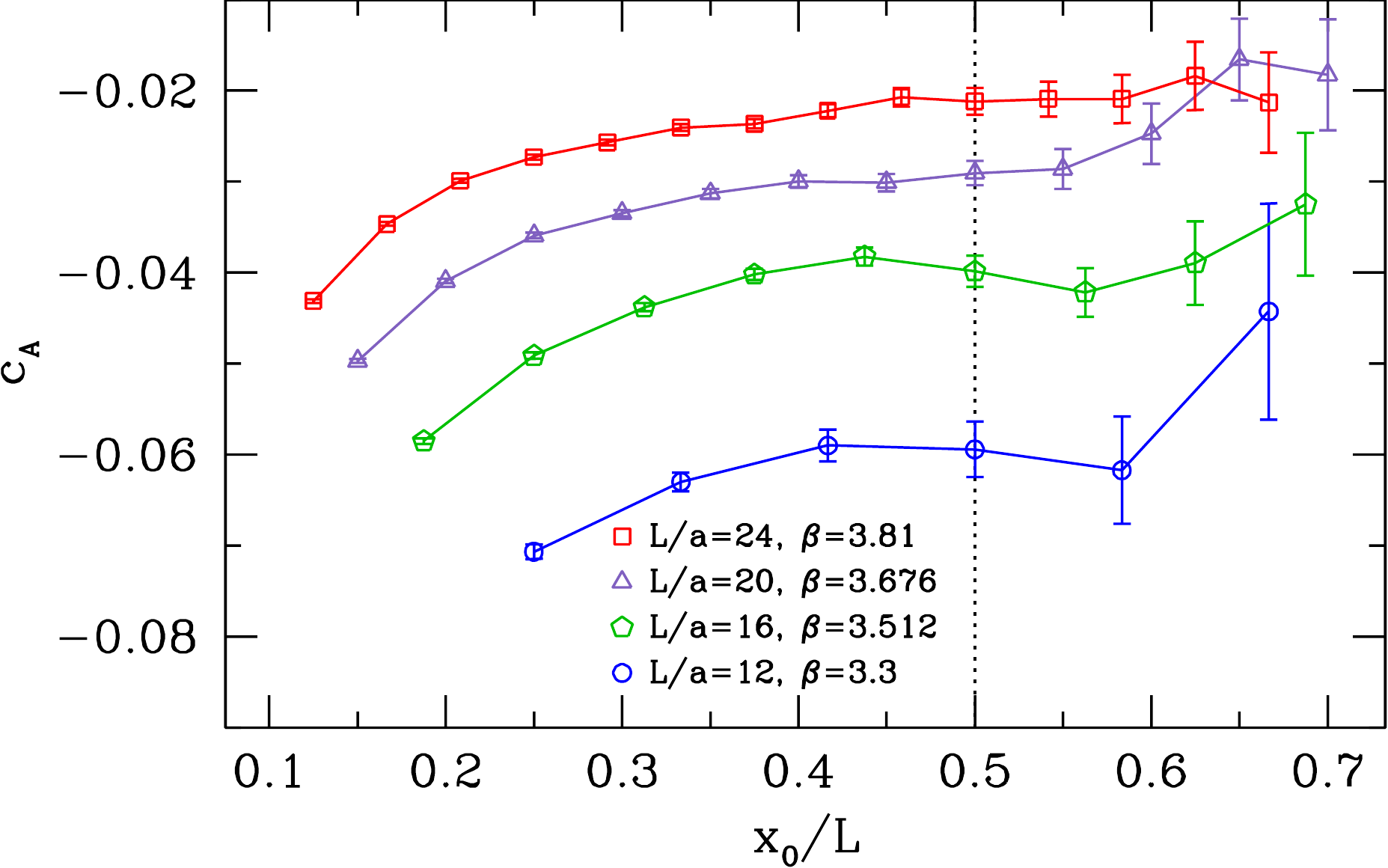}
\caption{The function $\ca(x_0)$ for all the ensembles used in the final 
analysis. The dotted vertical line at $x_0/L = 0.5$ indicates the 
space-time point used for our definition of $\ca$.}
\label{f:eff_ca}
\end{figure}
\begin{figure}
\centering
\includegraphics{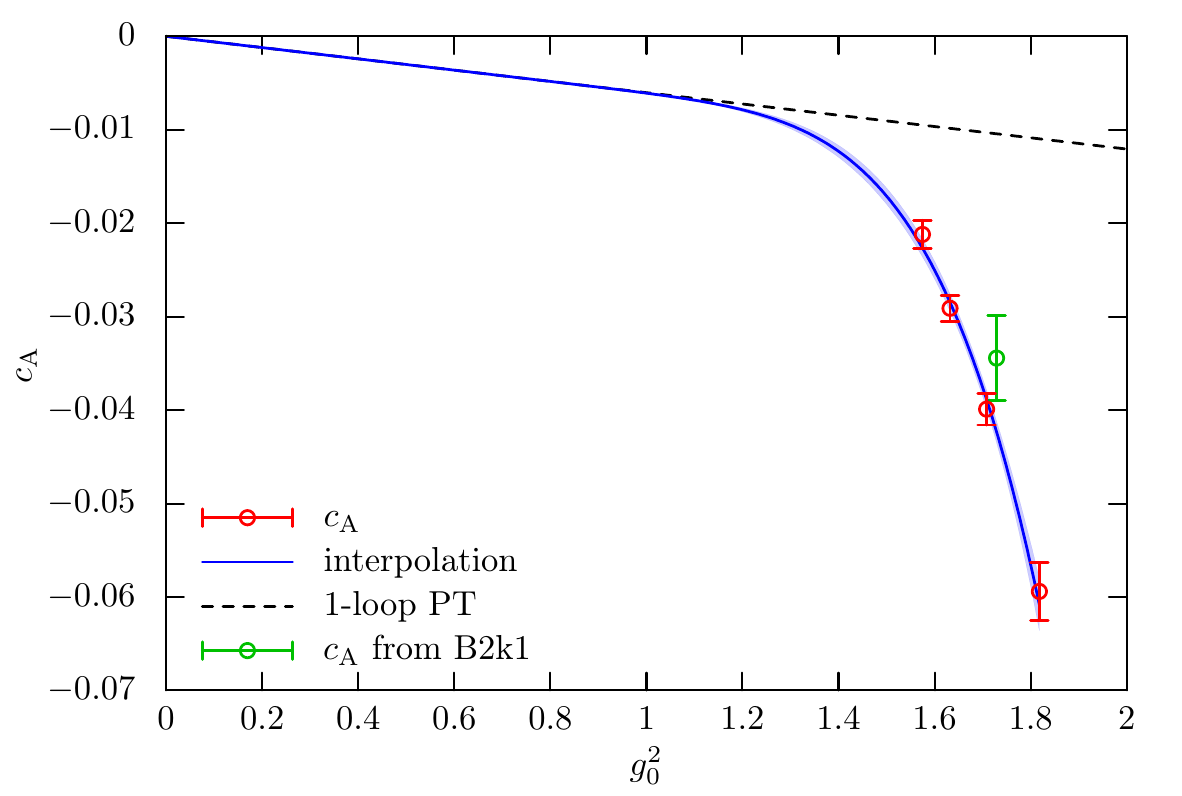}
\caption{Final results for \label{f:ca_plot}$\ca$ together with the
interpolation formula, Eqs.~(\ref{e:final}) and (\ref{e:par_final}).
The point from the B2k1 ensemble is not included in the fit. 
Note that our non-perturbative results fall appreciably apart from 1-loop 
perturbation theory.} 
\end{figure}

Finally, the function $\ca(x_0)$ is shown in Fig.~\ref{f:eff_ca} for the 
ensembles used in the analysis. It indicates that $\ca$ is rather 
independent of $x_0$ for points sufficiently distant from the boundary. 
Moreover, only a rather little variation of $\ca(x_0)$ is visible for 
$x_0\gtrsim 5a$; this reveals that high-energy states, 
which could induce large $\rmO(a)$ ambiguities in the improvement condition, 
are reasonably suppressed in this region.
As a compromise between cutoff effects and 
statistical errors, we take $x_0 = L/2$ as our final definition for $\ca$,
which besides is well within the regime where states with a distinct 
energy gap dominate the projected correlators.  
These data are plotted in Fig.~\ref{f:ca_plot}, together with an 
interpolation.

Before discussing these final results for $\ca$ at each bare coupling, we 
assess our systematic errors. In order to estimate the effect of our finite 
(but small) quark masses and thus of small violations of the constant quark 
mass condition on $\ca$, we study several different sets of ensembles, which
are identical apart from $|Lm_{\mathrm{PCAC}}|<0.3$ such as \{A1k1, A1k2\}, 
\{B1k1, B1k2, B1k3\}, and \{C1k1, C1k2, C1k3\}. Within each of these sets, 
the variation of the value of $\ca$ does not exceed more than about $1.5$ 
standard deviations. 
To quantify the cutoff effect, which results from  
$T = 3L/2 - a$, we compare \{A1k1, A1k2\} with A2k1, where the temporal 
extent is $T = 3L/2 +a$. We see that this results in a difference
which is significant at less than $1\sigma$ at our coarsest 
lattice spacing where it is largest.    
Lastly, we assess the error due to the deviation from our constant physics 
condition.
From Tab.~\ref{t:res} one infers
that the ensembles \{A1k1, B1k3, C1k3, D1k1\}, which enter into the
final analysis, have $\bar{g}_{\mathrm{GF},0}^2(L)$ between $\approx 14-18$, 
resulting in a $\approx 20\%$ variation assuming no cutoff effects. 
We can explicitly test the sensitivity of $\ca$ to this variation 
by means of the B2k1 ensemble, 
which differs with respect to the B1 ensembles
in $a$ by $\approx 6\%$ and in $\bar{g}_{\mathrm{GF},0}^2(L)$ by roughly the 
same amount as this overall $\bar{g}_{\mathrm{GF},0}^2(L)$-variation.
The value of $\ca$ determined on the B2k1 ensemble lies well within 
$1.5\sigma$ of the interpolating formula in Fig.~\ref{f:ca_plot}, 
so we are confident that this variation of $L$ does not result in a 
significant shift of $\ca$.  
Note that, after all, any imperfection in the constant physics condition of
this level and the systematic uncertainty in $\ca$ induced by it will only 
introduce $\Or(a^2)$ effects in quantities involving the improved axial 
current, which are negligible compared to other sources of errors and vanish 
in the continuum limit by definition. 

\begin{figure}[htb]
\centering
\includegraphics{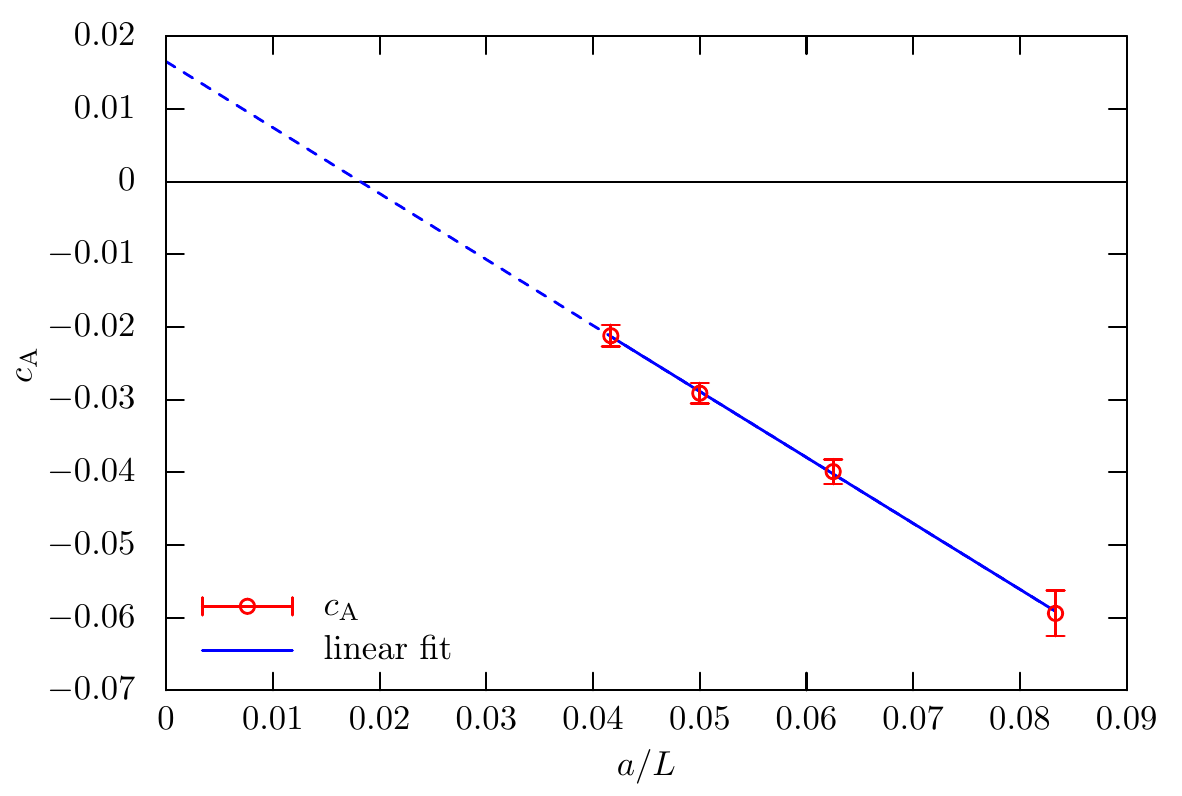}
\caption{\label{f:linear} $\ca$ versus the lattice spacing measured in
units of $L$ for our ensembles \{A1k1, B1k3, C1k3, D1k1\}.
A linear fit to the data within the region where we simulate is also shown.}
\end{figure}
We now move to the final results. 
In order to guide our choice for an interpolation formula, 
we observe that (for our choice of the improvement condition)
$\ca$ is almost linear in $a$ over the range of bare 
couplings which we simulate.
This approximate linearity is depicted in Fig.~\ref{f:linear}.
Notice that the linear behavior can \emph{not} extend all the way to $a=0$,
since this would be incompatible --- owing to the non-polynomial relation 
between $a$ and $g_0^2$ --- with a polynomial dependence of $\ca$ on
$g_0^2$ in the perturbative regime.
In our case, a naive linear fit to these data within the simulated region
does not even extrapolate to zero, which is the value predicted by
perturbation theory at tree-level: 
$\ca(g_0^2=0)=0$~\cite{pert:heatlie,impr:pap1}.
On the contrary, we interpret the behavior of the results according to 
$\ca=\mathrm{constant}+\mathrm{slope}\times a$ within the region of our
data such that the (non-vanishing) constant term removes the targeted
$\mathrm{O}(a)$ effects in the non-perturbative regime, while the
non-constant piece, describing the non-trivial dependence of $\ca$ on
$g_0^2$, only affects $\mathrm{O}(a^2)$ contributions in physical quantities.
Regardless of their actual size, these intrinsic $\Or(a)$ ambiguities
suggest to employ in the computation of physical quantities the
\emph{$g_0^2$-dependence} of $\ca$ induced by the constant physics
condition rather than, e.g., the constant piece alone, because one then
expects the remaining leading $\mathrm{O}(a^2)$ lattice artifacts to be
smaller and vanish uniformly in the $\Or(a)$ improved theory.

Motivated by the apparent linearity of $\ca$ as a function of the lattice
spacing within our data region, we choose the following interpolation 
formula for $\ca$ as a function of $g_0^2=6/\beta$:
\begin{align}\label{e:final}
  \ca(g_0^2) = -0.006033\,g_0^2 \times 
               \left[ 1 + \exp\left(p_0+\frac{p_1}{g_0^2}\right) \right], 
\end{align}
which is constrained to reproduce the 1-loop perturbative result of 
Ref.~\cite{impr:csw_iwa_pert} as $g_0^2$ goes to zero.   
The piece depending exponentially on the bare coupling is inspired by the
perturbative expression of the lattice spacing in terms of $g_0^2$, 
and it is inserted in order to capture the linear behavior reflected 
in Fig.~\ref{f:linear}.
Our non-perturbative results for $\ca$ are then compiled in 
Fig.~\ref{f:ca_plot}, together with this interpolating function with 
coefficients
\begin{align}\label{e:par_final}
p_0 = 9.2056, \quad p_1 = -13.9847,
\end{align}
which produces a $\chi^2/{\rm d.o.f.} \approx 0.85$;
they deviate markedly from the 1-loop perturbative prediction in the region
of our simulations.
Eq.~(\ref{e:final}), together with Eq.~(\ref{e:par_final}), represents our 
final result. This formula can be used at all bare couplings below 
$g_{0}^2 \approx 1.8$, together with the statistical errors on $\ca$
(i.e., on $c_{\mathrm{A},0}$ as given in Tab.~\ref{t:res}), which are 
$\approx 4-5\%$ near the largest simulated bare couplings and increase to 
$\approx 7-8\%$ near the smallest.

\section{Conclusions}\label{s:concl}

In this work we have determined non-perturbatively $\ca(g_0^2)$, the 
coefficient required for $\mathrm{O}(a)$~improvement of axial current 
matrix elements in lattice QCD with $\nf = 3$ flavors of Wilson quarks,
non-perturbative $\csw$~\cite{Bulava:2013cta} and the tree-level 
Symanzik-improved gauge action. 

The main result is the interpolation formula in Eqs.~(\ref{e:final})
and (\ref{e:par_final}), obtained using the standard improvement condition, 
which is imposed together with a variation of boundary wavefunctions in the 
PCAC relation with external states and along a line of constant physics in 
parameter space.
This implies that potentially large $\Or(a)$ ambiguities in $\ca$ are 
avoided and that its remaining intrinsic $\Or(a)$ ambiguities disappear 
smoothly toward the continuum limit.
Eq.~(\ref{e:final}) for $\ca(g_0^2)$ (with coefficients (\ref{e:par_final})) 
is valid for bare couplings below $g_{0}^2 \approx 1.8$
(or, equivalently, for lattice spacings $a \lesssim 0.09\,\Fm$).
We have treated the topology freezing, encountered in our simulations at the
finest lattice spacing (of around $0.045\,\Fm$), 
by restricting the improvement 
condition to the trivial topological sector for all ensembles considered. 

The gauge field ensembles entering this work were generated with $T = 3L/2$, 
in order to be re-used for the determination of the axial current 
renormalization constant $Z_{\rm A}$, 
see~\cite{lat14:caza_nf3} for a preliminary report.
After this is completed (the results of which will appear in a 
future publication), axial current matrix elements such as pseudoscalar 
meson decay constants can be calculated precisely from large-volume 
ensembles of gauge configurations at lattice spacings typically employed in
the context of phenomenological applications of lattice QCD.

\vskip 0.3cm

\noindent
{\bf Acknowledgements.}
We thank Rainer Sommer and Stefan Schaefer for helpful discussions, 
as well as Alberto Ramos who additionally provided an analysis 
program to measure the Wilson flow coupling. This work is supported by the 
grant {HE 4517/3-1} (J.~H.\ and C.~W.) of the Deutsche Forschungsgemeinschaft.
We gratefully acknowledge the computing time granted by the John von Neumann 
Institute for Computing (NIC) and provided on the supercomputer JUROPA at 
J\"{u}lich Supercomputing Centre (JSC).
Computer resources were also provided by DESY, Zeuthen (PAX cluster), 
the CERN `thqcd2' QCD HPC installation, and the ZIV of the University of 
M\"{u}nster (PALMA HPC cluster). 

\pagebreak


\begin{thebibliography}{10}

\bibitem{Wilson}
K.G. Wilson, {\it Confinement of quarks},
\newblock Phys. Rev. D10 (1974) 2445.

\bibitem{impr:Sym1}
K. Symanzik, {\it Continuum limit and improved action in lattice theories. 1.
  {P}rinciples and $\phi^4$ theory},
\newblock Nucl. Phys. B226 (1983) 187.

\bibitem{impr:Sym2}
K. Symanzik, {\it Continuum limit and improved action in lattice theories. 2.
  {O($N$)} nonlinear sigma model in perturbation theory},
\newblock Nucl. Phys. B226 (1983) 205.

\bibitem{impr:SW}
B. Sheikholeslami and R. Wohlert, {\it Improved continuum limit lattice action
  for {QCD} with {W}ilson fermions},
\newblock Nucl. Phys. B259 (1985) 572.

\bibitem{lambda:nf2}
P. Fritzsch et~al., {\it {The strange quark mass and Lambda parameter of two
  flavor QCD}},
\newblock Nucl. Phys. B865 (2012) 397,
  \href{http://xxx.lanl.gov/abs/1205.5380}{{\tt arXiv:1205.5380}}.

\bibitem{impr:ca_nf2}
M. Della~Morte, R. Hoffmann and R. Sommer, {\it Non-perturbative improvement of
  the axial current for dynamical {Wilson} fermions},
\newblock JHEP 03 (2005) 029,
  \href{http://xxx.lanl.gov/abs/hep-lat/0503003}{{\tt hep-lat/0503003}}.

\bibitem{Kaneko:2007wh}
T. Kaneko et~al., {\it {Non-perturbative improvement of the axial current with
  three dynamical flavors and the Iwasaki gauge action}},
\newblock JHEP 0704 (2007) 092,
  \href{http://xxx.lanl.gov/abs/hep-lat/0703006}{{\tt hep-lat/0703006}}.

\bibitem{Luscher:1984xn}
M. {L\"uscher} and P. Weisz, {\it {On-Shell Improved Lattice Gauge Theories}},
\newblock Commun. Math. Phys. 97 (1985) 59.

\bibitem{silvia:universality}
S. Necco, {\it {Universality and scaling behavior of RG gauge actions}},
\newblock Nucl. Phys. B683 (2004) 137,
  \href{http://xxx.lanl.gov/abs/hep-lat/0309017}{{\tt hep-lat/0309017}}.

\bibitem{Bruno:2014jqa}
M. Bruno et~al., {\it {Simulation of QCD with $N_{\rm f}=2+1$ flavors of
  non-perturbatively improved Wilson fermions}},
\newblock JHEP 1502 (2015) 043, \href{http://xxx.lanl.gov/abs/1411.3982}{{\tt
  arXiv:1411.3982}}.

\bibitem{Bulava:2013cta}
J. Bulava and S. Schaefer, {\it {Improvement of $N_{\rm f}=3$ lattice QCD with
  Wilson fermions and tree-level improved gauge action}},
\newblock Nucl. Phys. B874 (2013) 188,
  \href{http://xxx.lanl.gov/abs/1304.7093}{{\tt arXiv:1304.7093}}.

\bibitem{impr:csw_iwa_pert}
S. Aoki, R. Frezzotti and P. Weisz, {\it Computation of the improvement
  coefficient $c_{\mathrm{sw}}$ to 1-loop with improved gluon actions},
\newblock Nucl. Phys. B540 (1999) 501,
  \href{http://xxx.lanl.gov/abs/hep-lat/9808007}{{\tt hep-lat/9808007}}.

\bibitem{impr:pap3}
M. {L\"uscher} et~al., {\it Non-perturbative {O($a$)} improvement of lattice
  {QCD}},
\newblock Nucl. Phys. B491 (1997) 323,
  \href{http://xxx.lanl.gov/abs/hep-lat/9609035}{{\tt hep-lat/9609035}}.

\bibitem{SF:LNWW}
M. {L\"uscher}, R. Narayanan, P. Weisz and U. Wolff, {\it {The Schr\"odinger
  functional: A Renormalizable probe for non-Abelian gauge theories}},
\newblock Nucl. Phys. B384 (1992) 168,
  \href{http://xxx.lanl.gov/abs/hep-lat/9207009}{{\tt hep-lat/9207009}}.

\bibitem{SF:stefan1}
S. Sint, {\it On the {Schr\"odinger} functional in {QCD}},
\newblock Nucl. Phys. B421 (1994) 135,
  \href{http://xxx.lanl.gov/abs/hep-lat/9312079}{{\tt hep-lat/9312079}}.

\bibitem{Luscher:2014kea}
M. {L\"uscher}, {\it {Step scaling and the Yang-Mills gradient flow}},
\newblock JHEP 1406 (2014) 105, \href{http://xxx.lanl.gov/abs/1404.5930}{{\tt
  arXiv:1404.5930}}.

\bibitem{DellaMorte:2008xb}
M. Della~Morte, R. Sommer and S. Takeda, {\it {On cutoff effects in lattice QCD
  from short to long distances}},
\newblock Phys. Lett. B672 (2009) 407,
  \href{http://xxx.lanl.gov/abs/0807.1120}{{\tt arXiv:0807.1120}}.

\bibitem{lat14:caza_nf3}
J. Bulava, M. {Della Morte}, J. Heitger and C. Wittemeier, {\it
  {Non-perturbative improvement and renormalization of the axial current in
  $N_{\rm f}=3$ lattice QCD}},
\newblock PoS LATTICE2014 (2014) 283.

\bibitem{lat13:felix}
P. Fritzsch, A. Ramos and F. Stollenwerk, {\it {Critical slowing down and the
  gradient flow coupling in the Schr\"odinger functional}},
\newblock PoS LATTICE2013 (2013) 461,
  \href{http://xxx.lanl.gov/abs/1311.7304}{{\tt arXiv:1311.7304}}.

\bibitem{flow:ML}
M. {L\"uscher}, {\it {Properties and uses of the Wilson flow in lattice QCD}},
\newblock JHEP 1008 (2010) 071, \href{http://xxx.lanl.gov/abs/1006.4518}{{\tt
  arXiv:1006.4518}}.

\bibitem{flow:LW}
M. {L\"{u}scher} and P. Weisz, {\it {Perturbative analysis of the gradient flow
  in non-abelian gauge theories}},
\newblock JHEP 1102 (2011) 051, \href{http://xxx.lanl.gov/abs/1101.0963}{{\tt
  arXiv:1101.0963}}.

\bibitem{algo:openQCD}
M. {L\"uscher} and S. Schaefer, {\it {Lattice QCD with open boundary conditions
  and twisted-mass reweighting}},
\newblock Comput. Phys. Commun. 184 (2013) 519,
  \href{http://xxx.lanl.gov/abs/1206.2809}{{\tt arXiv:1206.2809}}.

\bibitem{impr:babp_nf2}
P. Fritzsch, J. Heitger and N. Tantalo, {\it {Non-perturbative improvement of
  quark mass renormalization in two-flavour lattice QCD}},
\newblock JHEP 08 (2010) 074, \href{http://xxx.lanl.gov/abs/1004.3978}{{\tt
  arXiv:1004.3978}}.

\bibitem{nara:rainer}
R. Sommer, {\it Non-perturbative {QCD}: renormalization, {O}($a$)-improvement
  and matching to {H}eavy {Q}uark {E}ffective {T}heory},
\newblock Lectures given at the {\it ILFTN Workshop on ``Perspectives in
  Lattice QCD''}, Nara, Japan, 31 October -- 11 November 2005  (World
  Scientific 2008), \href{http://xxx.lanl.gov/abs/hep-lat/0611020}{{\tt
  hep-lat/0611020}}.

\bibitem{flow:FR}
P. Fritzsch and A. Ramos, {\it {The gradient flow coupling in the
  Schr\"{o}dinger Functional}},
\newblock JHEP 1310 (2013) 008, \href{http://xxx.lanl.gov/abs/1301.4388}{{\tt
  arXiv:1301.4388}}.

\bibitem{algo:RHMC}
M. Clark and A. Kennedy, {\it {Accelerating dynamical fermion computations
  using the rational hybrid Monte Carlo (RHMC) algorithm with multiple
  pseudofermion fields}},
\newblock Phys. Rev. Lett. 98 (2007) 051601,
  \href{http://xxx.lanl.gov/abs/hep-lat/0608015}{{\tt hep-lat/0608015}}.

\bibitem{Hasenbusch:2002ai}
M. Hasenbusch and K. Jansen, {\it Speeding up lattice {QCD} simulations with
  clover-improved {W}ilson fermions},
\newblock Nucl. Phys. B659 (2003) 299,
  \href{http://xxx.lanl.gov/abs/hep-lat/0211042}{{\tt hep-lat/0211042}}.

\bibitem{Sexton:1992nu}
J.C. Sexton and D.H. Weingarten, {\it {Hamiltonian evolution for the hybrid
  Monte Carlo algorithm}},
\newblock Nucl. Phys. B380 (1992) 665.

\bibitem{Omelyan}
I.M. Omelyan, I.P. Mryglod and R. Folk, {\it {Symplectic analytically
  integrable decomposition algorithms: classification, derivation, and
  application to molecular dynamics, quantum and celestial mechanics
  simulations}},
\newblock Comp. Phys. Comm. 151 (2003) 272.

\bibitem{lat08:filippo}
M. {L\"uscher} and F. Palombi, {\it {Fluctuations and reweighting of the quark
  determinant on large lattices}},
\newblock PoS LATTICE2008 (2008) 049,
  \href{http://xxx.lanl.gov/abs/0810.0946}{{\tt arXiv:0810.0946}}.

\bibitem{Luscher:2003qa}
M. {L\"uscher}, {\it {Solution of the Dirac equation in lattice QCD using a
  domain decomposition method}},
\newblock Comput. Phys. Commun. 156 (2004) 209,
  \href{http://xxx.lanl.gov/abs/hep-lat/0310048}{{\tt hep-lat/0310048}}.

\bibitem{Bruno:2014ova}
M. Bruno, S. Schaefer and R. Sommer, {\it {Topological susceptibility and the
  sampling of field space in $N_{\rm f}=2$ lattice QCD simulations}},
\newblock JHEP 1408 (2014) 150, \href{http://xxx.lanl.gov/abs/1406.5363}{{\tt
  arXiv:1406.5363}}.

\bibitem{Leutwyler:1992yt}
H. Leutwyler and A.V. Smilga, {\it {Spectrum of Dirac operator and role of
  winding number in QCD}},
\newblock Phys. Rev. D46 (1992) 5607.

\bibitem{Wolff:2003sm}
U. Wolff, {\it {Monte Carlo} errors with less errors},
\newblock Comput. Phys. Commun. 156 (2004) 143,
  \href{http://xxx.lanl.gov/abs/hep-lat/0306017}{{\tt hep-lat/0306017}}.

\bibitem{algo:csd}
S. Schaefer, R. Sommer and F. Virotta, {\it {Critical slowing down and error
  analysis in lattice QCD simulations}},
\newblock Nucl. Phys. B845 (2011) 93,
  \href{http://xxx.lanl.gov/abs/1009.5228}{{\tt arXiv:1009.5228}}.

\bibitem{pert:heatlie}
G. Heatlie et~al., {\it The improvement of hadronic matrix elements in lattice
  {QCD}},
\newblock Nucl. Phys. B352 (1991) 266.

\bibitem{impr:pap1}
M. {L\"uscher}, S. Sint, R. Sommer and P. Weisz, {\it Chiral symmetry and
  {O($a$)} improvement in lattice {QCD}},
\newblock Nucl. Phys. B478 (1996) 365,
  \href{http://xxx.lanl.gov/abs/hep-lat/9605038}{{\tt hep-lat/9605038}}.

\end{thebibliography}

\end{document}